# Thermal radiation, radiation force and dynamics of a polarizable particle in vacuum

G.V. Dedkov and A.A. Kyasov

Kabardino-Balkarian State University, Nanoscale Physics Group, Nalchik, Russian Federation

We discuss the basic expressions and interrelations between various physical quantities describing the fluctuation-electromagnetic interaction of a small polarizable particle during relativistic motion relative to the blackbody radiation, namely the radiation tangential force, rate of heating, intensity of thermal radiation/absorption, and acceleration. We also obtain an explicit formula for the frictional force acting on the particle in its rest frame and discuss its connection with tangential force in the reference frame of background radiation.

1. **Introduction**

The problem of blackbody friction – the thermal drag force on a particle moving through a thermal bath at velocity $V$, has attracted attention since pioneering work by Einstein and Hopf [1]. But nearly a century passed until several authors [2, 3] obtained the formulas for nonrelativistic drag force on a small polarizable particle. The next step was made in [4, 5]. Using the formalism of fluctuation-electromagnetic theory, we obtained the relativistic expression for the radiation force assuming that the particle and electromagnetic field are characterized by local equilibrium temperatures. Further on, the basic formula [4, 5] was recovered using a fully covariant framework [6, 7]. In [8], these results were used to calculate the intensity of thermal radiation of relativistic polarizable particle. Closely related problems concerning the radiation of rotating particles [9–11] and Cherenkov frictional losses [12] were also in the focus of recent investigations.

It is clear that blackbody friction leads to the energy dissipation, heating and radiation of a moving particle. On the other hand, the temperature imbalance causes the thermal radiation even when the particle is at rest. For a moving particle the situation is more complicated, since the thermal radiation (absorption) affects the rest energy of the particle and its dynamics. This fact was first noted by Polevoi [13] in the problem of friction between the plates moving relative one another. Therefore, all aforementioned physical quantities are not independent one another. This work aims at discussing these issues in more detail and obtaining new important links between the intensity of thermal radiation, radiation force, dynamics and acceleration of a polarizable particle.

The radiation force in the reference frame related to the particle ($\Sigma'$) and in the reference frame of background radiation ($\Sigma$) are interrelated. Some authors have used this fact to calculate the radiation force in $\Sigma$ using the corresponding expression in $\Sigma'$, but obtained different results (see, for example, [14, 15] and references therein). In a recent paper [15] the history of the problem and controversies in the results of various authors relating the radiation force in reference systems $\Sigma$ and $\Sigma'$ were discussed and some points of misalignment were featured. However, to get a more complete description and for self-consistency of the problem solution these points have to be clarified in more detail. This is another aim of the present work.

## 2. Theory

Following [4, 5], consider a small particle of radius $R$ uniformly moving with a velocity $V$ through thermalized photonic gas with temperature $T_2$ (Fig. 1). Let the surface $\sigma$ encircles the particle at a large enough distance so that the fluctuation-electromagnetic field on $\sigma$ represents the wave field. The reference frames $\Sigma$ and $\Sigma'$ correspond to the reference frames of background radiation and particle, respectively. Initially, the particle has the temperature $T_1$ in its own reference frame. We also assume that $R << \min(2\pi\hbar c/k_B T_1, 2\pi\hbar c/k_B T_2)$. In this case, the particle can be considered as a point-like dipole with fluctuating dipole and magnetic moments $\mathbf{d}(t), \mathbf{m}(t)$, and its material properties are described by the frequency-dependent dielectric and (or) magnetic polarizabilities $\alpha_e(\omega)$, $\alpha_m(\omega)$.

According to the conventional form of the energy conservation law of the electromagnetic field within the volume $\Omega$ restricted by surface $\sigma$ one may write [8]

$$-\frac{dW}{dt} = \oint_\sigma \mathbf{S} \cdot d\vec{\sigma} + \int_\Omega \langle \mathbf{j} \cdot \mathbf{E} \rangle d^3r,$$

$$W = \int_\Omega \frac{\langle \mathbf{E}^2 \rangle + \langle \mathbf{H}^2 \rangle}{8\pi} d^3r, \qquad (2)$$

$$\mathbf{S} = \frac{c}{4\pi} \langle \mathbf{E} \times \mathbf{H} \rangle, \qquad (3)$$

where $W$ is the energy of the fluctuating electromagnetic field in the volume $\Omega$ and $\mathbf{S}$ the Poynting vector of this field. The angular brackets in (1)—(3) denote total quantum and statistical averaging. In quasistationary approximation, $dW/dt = 0$, from (1) it follows where $W$ is the energy of the fluctuating electromagnetic field in the volume $\Omega$ and $\mathbf{S}$ the Poynting vector of this field. The angular brackets in (1)—(3) denote total quantum and statistical averaging. In quasistationary approximation, $dW/dt = 0$, from (1) it follows

$$I = \oint_\sigma \mathbf{S}\cdot d\vec{\sigma} = -\int_\Omega \langle \mathbf{j}\cdot\mathbf{E}\rangle d^3 r \equiv I_1 - I_2, \tag{4}$$

where $I_1 = I_1(T_1)$ describes the intensity of thermal radiation emitted by the particle in vacuum, and $I_2 = I_2(T_2)$ describes the intensity of absorbed background radiation.

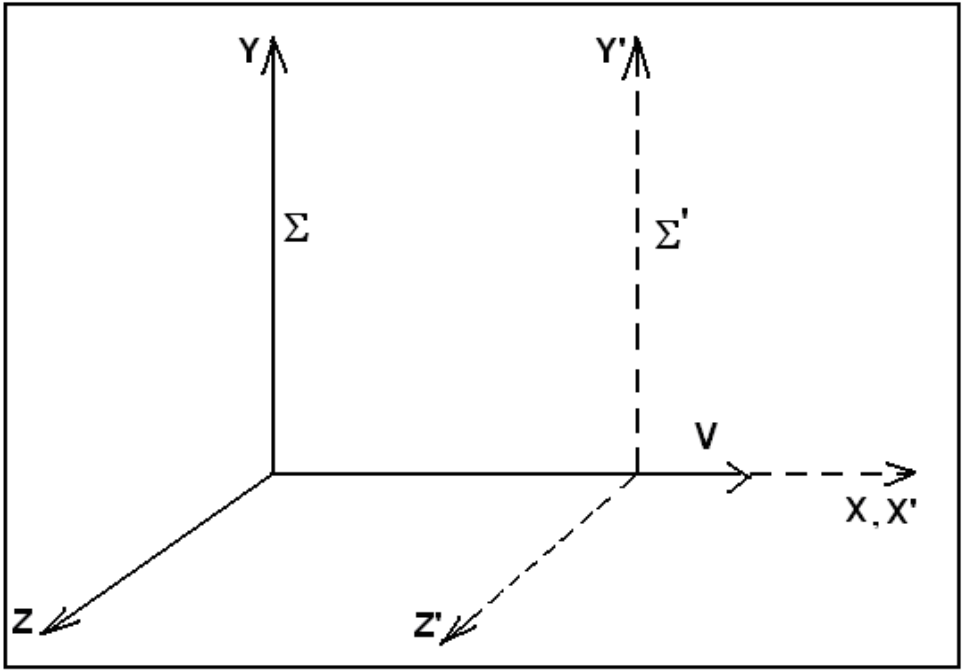

Fig. 1. Geometrical configuration and the two frames of reference considered. $\Sigma$ is the rest frame of radiation; $\Sigma'$ moves along $+x'$ with speed $V$.

We now recall the relations between the dissipation integral $\int_\Omega \langle \mathbf{j}\cdot\mathbf{E}\rangle d^3 r$ and the Lorentz force $\mathbf{F}$ expressed in terms of the fluctuating dipole moments $\mathbf{d}(t), \mathbf{m}(t)$ [16]

$$\int_\Omega \langle \mathbf{j}\cdot\mathbf{E}\rangle d^3 r - F_x \cdot V = \langle \dot{\mathbf{d}}\cdot\mathbf{E} + \dot{\mathbf{m}}\cdot\mathbf{H}\rangle, \tag{5}$$

$$\mathbf{F} = \int \langle \rho\mathbf{E}\rangle d^3 r + \frac{1}{c}\int \langle \mathbf{j}\times\mathbf{H}\rangle d^3 r = \langle \nabla(\mathbf{d}\cdot\mathbf{E} + \mathbf{m}\cdot\mathbf{H})\rangle, \tag{6}$$

where $F_x$ in (5) is the projection of $\mathbf{F}$ onto the velocity direction and the dots denote the derivatives with respect to time. It should be noted, that the fields $\mathbf{E}$ and $\mathbf{H}$ in the right hand sides of (5)–(7) should be taken at the point of particle location. Denoting

$$dQ/dt = \langle \mathbf{d} \cdot \mathbf{E} + \dot{\mathbf{m}} \cdot \mathbf{H} \rangle, \qquad (7)$$

from (4)–(7) one obtains

$$I = I_1 - I_2 = -\left(\frac{dQ}{dt} + F_x V\right). \qquad (8)$$

Despite its simplicity, Eq. (8) was used in the problem of thermal radiation of a moving particle only recently [8].

To proceed further, it is convenient to write the energy conservation law (1) in another form

$$-\frac{d}{dt}\left(W + \frac{mc^2}{\sqrt{1-\beta^2}}\right) = I_1 - I_2 = I, \qquad (9)$$

where $\beta = V/c$ and $m$ is the particle mass. Using (9), with allowance for (4) and the condition of quasistationarity $dW/dt = 0$, one obtains ($\gamma = 1/\sqrt{1-\beta^2}$)

$$-\gamma^3 mV \frac{dV}{dt} - \gamma \frac{dm}{dt} c^2 = I. \qquad (10)$$

Furthermore, writing down the dynamics equation in reference frame $\Sigma$

$$\frac{d}{dt}\left(\frac{mV}{\sqrt{1-V^2/c^2}}\right) = F_x, \qquad (11)$$

we obtain

$$\gamma^3 m \frac{dV}{dt} + \gamma V \frac{dm}{dt} = F_x. \qquad (12)$$

By solving (10) and (12) with respect to $dV/dt$ and $dm/dt$ with allowance for (8) one obtains

$$\gamma^3 m \frac{dV}{dt} = F_x - \frac{\beta}{1-\beta^2} \frac{\dot{Q}}{c}, \tag{13}$$

$$\frac{dm}{dt} = \frac{\gamma \dot{Q}}{c^2}. \tag{14}$$

It is worth noting that equations similar to (8), (13) and (14) were obtained by Polevoi [13] in the problem of friction between two thick plates in relative motion.

Now we consider the rate of heating of a moving particle and recall the general equation [16]

$$\int_{\Omega'} \langle \mathbf{j}' \cdot \mathbf{E}' \rangle d^3 r' = \gamma^2 \left( \int_{\Omega} \langle \mathbf{j} \cdot \mathbf{E} \rangle d^3 r - F_x \cdot V \right) = \gamma^2 \dot{Q}, \tag{15}$$

where the primed quantities correspond to reference frame $\Sigma'$ (Fig. 1). Equation (15) can be verified by performing the Lorentz transformations of the incoming quantities taken in coordinate systems $\Sigma$ and $\Sigma'$. The left part of (15) is associated with the particle heating rate

$$\frac{dQ'}{dt'} = \frac{d}{dt'}(C_s m T_1) = \gamma \frac{d}{dt}(C_s m T_1), \tag{16}$$

where $C_s$ is the specific heat. By neglecting the change of $C_s$ during the time of heating, from (14)–(16) it follows

$$C_s m \frac{dT_1}{dt} = \gamma \dot{Q} \left(1 - \frac{C_s T_1}{c^2}\right). \tag{17}$$

Since the condition $C_s T_1 / c^2 \ll 1$ holds even at the melting point of solids, Eq. (17) reduces to

$$C_s m \frac{dT_1}{dt} = \gamma \dot{Q}. \tag{18}$$

It is this equation was used in [8] when calculating the particle heating. Therefore, the change in the particle rest mass is negligible when calculating the temperature evolution, in contrast to the dynamics equation (13).

### 3. Radiation force and intensity of radiation

The right-hand sides of (6) and (7) are calculated in line with [4, 5] by representing them as the products of spontaneous and induced quantities and calculating the correlators with the help of relativistic fluctuation-dissipation relations. The final results have the form

$$F_x = -\frac{\hbar\gamma}{\pi c^4}\int_0^\infty d\omega\,\omega^4 \int_{-1}^1 dx\,x(1+\beta x)^2 \alpha''(\omega_\beta) \cdot$$
$$\cdot [\coth(\hbar\omega/2k_B T_2) - \coth(\hbar\omega_\beta/2k_B T_1)] \quad (19)$$

$$\frac{dQ}{dt} = \frac{\hbar\gamma}{\pi c^3}\int_0^\infty d\omega\,\omega^4 \int_{-1}^1 dx\,(1+\beta x)^3 \alpha''(\omega_\beta) \cdot$$
$$\cdot [\coth(\hbar\omega/2k_B T_2) - \coth(\hbar\omega_\beta/2k_B T_1)] \quad (20)$$

where $\omega_\beta = \gamma\omega(1+\beta x)$ and $\alpha''$ denotes the imaginary part of the sum of the electric and magnetic polarizability. The terms in (19), (20) which depend on $T_1$ correspond to the contributions related with spontaneous fluctuations of the particle dipole moments, whereas the terms depending on $T_2$ correspond to the contributions related with fluctuation fields of the vacuum background. Substituting (19), (20) into (8) yields [8]

$$I = -\frac{2\hbar\gamma}{\pi c^3}\int_0^\infty d\omega\,\omega^4 \int_{-1}^1 dx\,(1+\beta x)^2 \alpha''(\omega_\beta) \cdot$$
$$\cdot \left[\frac{1}{\exp(\hbar\omega/k_B T_2)-1} - \frac{1}{\exp(\hbar\omega_\beta)/k_B T_1)-1}\right]. \quad (21)$$

The first and second terms in (21) describe the absorbed intensity of background radiation and the intensity of radiation emitted by the particle.

### 4. Particle acceleration

Substituting (19), (20) into the right-hand side of (13), one obtains

$$\left(F_x - \frac{\beta}{1-\beta^2}\frac{dQ/dt}{c}\right) =$$
$$= -\frac{2\hbar\gamma^3}{\pi c^4}\int_0^\infty d\omega\,\omega^4 \int_{-1}^1 dx\,\frac{\alpha''(\omega_\beta)(x+\beta)(1+\beta x)^2}{(\exp(\hbar\omega/k_B T_2)-1)} - J_1(T_1) + J_2(T_1) \quad (22)$$

where the $T_1$-dependent terms $J_1(T_1)$ and $J_2(T_1)$ cancel each other (see Appendix)

According to (13) and (22), the acceleration $d\beta/dt$ is described by the equation

$$mc\frac{d\beta}{dt} = -\frac{2\hbar}{\pi c^4}\int_0^\infty d\omega\, \omega^4 \int_{-1}^{1} dx\, \frac{\alpha''[\gamma\omega(1+\beta x)](x+\beta)(1+\beta x)^2}{(\exp(\hbar\omega/k_B T_2)-1)}. \tag{23}$$

From (23) it follows that the particle acceleration is always negative and depends only on the background temperature $T_2$, $\beta$ and material properties. This is in agreement with [14, 15]. It is worthwhile noting that the time scale of particle deceleration (according to (23)) is much longer than the time scale of particle heating/cooling (according to (17) and (18)). Due to this, the kinetics of particle heating and can be studied at fixed values of the velocity factor.

**5. Friction force in the reference frame of particle**

The relation between the radiation force $F_x$ (19) in $\Sigma$ and the corresponding force $F'_x$ in $\Sigma'$ is simply obtained by differentiating the Lorentz transformation of the particle momentum in reference frames $\Sigma$ and $\Sigma'$, $p_x = \gamma(p'_x + Vm)$:

$$F_x = F'_x + V\frac{dm}{dt'} = F_x + \gamma V\frac{dm}{dt}, \tag{24}$$

Substituting (14) into (24) yields

$$F'_x = F_x - \frac{\beta}{1-\beta^2}\frac{dQ/dt}{c}. \tag{25}$$

Comparing (25) and (13) one can see that Eq. (13) can be written in the form

$$mc\frac{d\beta}{dt} = (1-\beta^2)^{3/2} F'_x, \tag{26}$$

where $F'_x$ is given by the first term in (22), namely

$$F'_x = -\frac{2\hbar\gamma^3}{\pi c^4}\int_0^\infty d\omega\, \omega^4 \int_{-1}^{1} dx\, \frac{\alpha''(\gamma\omega(1+\beta x))(x+\beta)(1+\beta x)^2}{(\exp(\hbar\omega/k_B T_2)-1)}. \tag{27}$$

Further on, by transforming the double integral (27) one obtains the following expression

$$F'_x = \frac{\hbar}{\pi c^4} \int_0^\infty d\omega \omega^4 \int_{-1}^1 dx\, x\, \alpha''(\omega) \coth\left(\frac{\hbar \omega \gamma (1+\beta x)}{2k_B T_2}\right) \tag{28}$$

Equations (13), (14), (17)–(23) and (28) summarize the main results of this work, providing closed description of radiation, dynamics and kinetics of heating of a moving particle with respect to the c. m. of background radiation, as well as their link to the results obtained in the particle rest frame.

Formula (28) coincides with the corresponding result in [15]. From (26) and (28) it follows that the particle is decelerated in reference frame $\Sigma$ and the acceleration is determined by the friction force $F'_x$ which acts in reference frame $\Sigma'$.

In the limit $V \ll c$ Eq. (28) reduces to [3]

$$F'_x = -\frac{\hbar^2 V}{3\pi c^5 k_B T_2} \int_0^\infty d\omega \omega^5 \frac{\alpha''(\omega)}{\sinh^2(\hbar\omega/2k_B T_2)} \;. \tag{29}$$

## 6. Discussion

The relativistic theory of black-body friction developed in [14, 15] is based on the calculation of the friction force $F'_x$ corresponding to the reference frame of particle. After some attempts in calculating the blackbody friction the author [15] has come to the conclusion that his resultant formula for $F_x$ agrees with our results [4, 5] and with [6, 7]. In our opinion, at the initial stage of consideration, it is the use of different frames of reference in calculating the blackbody friction has led to some puzzles and incorrect results in expressions of the friction force $F_x$ which do not include the contributions of particle radiation with temperature $T_1$. This point was criticized by us in [5], where we noted that the force $F_x$ should include both the contribution of background radiation with temperature $T_2$ and the contribution of particle radiation with temperature $T_1$.

In our works [4, 5] we did not calculate the force $F'_x$ and our criticism of the results [14] and preceding results of these authors was directed to the incorrect result for $F_x$ in reference frame $\Sigma$, depending only on $T_2$. At the same time, the accurate formula for $F_x$ (see (19)) contains the contributions both $T_1$ and $T_2$. Moreover, in view of the results obtained in this work, in order to reach self-consistency of the problem solution one needs to know correct expressions for $F_x$, $F'_x$ and their interrelation with other important physical quantities.

Some of our discussion comments made in [5] relating to the results [14] were attempted to explain the controversies between our results and those in [14], and they were not sufficiently accurate. However, this could not be the reason for the doubts in correctness of the basic results [4, 5]. This work, as we believe, resolves the main contradictions and highlights the essence of the problem much more clear. Therefore, the criticism of our results [4, 5] in [15] seems to be groundless.

## 7. Summary

We have obtained a self-consistent system of equations describing the dynamics, heating and thermal radiation of a polarizable particle with arbitrary temperature moving relative to the background radiation. Several new important interrelations have been found. We also obtained an explicit expression for the friction force in the rest frame of a particle and its link with the radiation force determined in the reference frame of background radiation. The results can be used for studying the noncontact friction, dynamics and radiation of particles in cavities, and interaction of the cosmic dust matter with radiation.

**Appendix**

Consider the integrals $J_1(T_1)$ and $J_2(T_1)$ in (20).

$$J_1(T_1) = \frac{2\hbar\gamma}{\pi c^4} \int_0^\infty d\omega\, \omega^4 \int_{-1}^1 dx\, \frac{x(1+\beta x)^2 \alpha''(\gamma\omega(1+\beta x))}{(\exp(\hbar\gamma\omega(1+\beta x)/k_B T_1) - 1)} \tag{A1}$$

$$J_2(T_1) = \frac{2\hbar\gamma^3\beta}{\pi c^4} \int_0^\infty d\omega\, \omega^4 \int_{-1}^1 dx\, \frac{(1+\beta x)^3 \alpha''(\gamma\omega(1+\beta x))}{(\exp(\hbar\gamma\omega(1+\beta x)/k_B T_1) - 1)} \tag{A2}$$

Introducing new variable $\omega' = \gamma\omega(1+\beta x)$, equations (A1), (A2) take the form

$$J_1(T_1) = \frac{2\hbar\gamma^{-4}}{\pi c^4} \int_0^\infty d\omega'\, \omega'^4 \frac{\alpha''(\omega')}{[\exp(\hbar\omega'/k_B T_1) - 1]} \int_{-1}^1 dx\, \frac{x}{(1+\beta x)^3} \tag{A3}$$

$$J_2(T_1) = \frac{2\hbar\gamma^{-2}\beta}{\pi c^4} \int_0^\infty d\omega'\, \omega'^4 \frac{\alpha''(\omega')}{[\exp(\hbar\omega'/k_B T_1) - 1]} \int_{-1}^1 dx\, \frac{1}{(1+\beta x)^2} \tag{A4}$$

The inner integrals in (A3) and (A4) are equal to $-2\beta\gamma^4$ and $2\gamma^2$ and, therefore, $-J_1(T_1) + J_2(T_1) = 0$.